\documentclass[twocolumn, aps, prb, showpacs, showkeys]{revtex4}
\usepackage{hyperref}
\usepackage{amsmath}
\usepackage{amssymb}
\usepackage{units}
\usepackage{graphicx}
\usepackage{psfrag}
\usepackage{tabularx}
\usepackage{booktabs}
\usepackage{bigstrut}

\begin{document}

\title{Quasiparticle GW calculations for solids, molecules and 2D materials}
\date{\today}
\author{Falco H\"user}
\email{falco.hueser@fysik.dtu.dk}
\affiliation{Center for Atomic-scale Materials Design (CAMD), Department of 
Physics\\
Technical University of Denmark, 2800 Kgs. Lyngby, Denmark}
\author{Thomas Olsen}
\affiliation{Center for Atomic-scale Materials Design (CAMD), Department of 
Physics\\
Technical University of Denmark, 2800 Kgs. Lyngby, Denmark}
\author{Kristian S. Thygesen}
\affiliation{Center for Atomic-scale Materials Design (CAMD), Department of 
Physics\\
Technical University of Denmark, 2800 Kgs. Lyngby, Denmark}
\affiliation{Center for Nanostructured Graphene (CNG)\\
Technical University of Denmark, 2800 Kgs. Lyngby, Denmark}

\begin{abstract}
  We present a plane wave implementation of the G$_0$W$_0$
  approximation within the projector augmented wave method code GPAW. The
  computed band gaps of ten bulk semiconductors and insulators deviate
  on average by $\unit[0.2]{eV}$ ($\sim 5 \%$) from the experimental values --
  the only exception being ZnO where the calculated band gap is around
  $\unit[1]{eV}$ too low. Similar relative deviations are found for
  the ionization potentials of a test set of 32 small molecules. The
  importance of substrate screening for a correct description of
  quasiparticle energies and Fermi velocities in supported 2D
  materials is illustrated by the case of graphene/\textit{h}-BN
  interfaces. Due to the long range Coulomb interaction between
  periodically repeated images, the use of a truncated interaction is
  found to be essential for obtaining converged results for 2D
  materials. For all systems studied, a plasmon pole approximation is
  found to reproduce the full frequency results to within
  $\unit[0.2]{eV}$ with a significant gain in computational speed. As 
  alternative to G$_0$W$_0$, the efficient local GLLBSC potential yields 
  significantly better results than the PBE0 hybrid.
  For completeness, we provide a mathematically rigorous and physically
  transparent introduction to the notion of quasiparticle states.
\end{abstract}


\maketitle

\section{Introduction}

For several decades, density functional theory (DFT)
\cite{Hohenberg-Kohn_PhysRev1964, Kohn-Sham_PhysRev1965} has been the
method of choice for electronic structure calculations due to its
unique compromise between accuracy and efficiency. Large efforts have
been made to develop better exchange-correlation (xc-) functionals
continuously pushing the quality of total energy calculations towards
the limit of chemical accuracy.  However, it is well known that the
Kohn-Sham single-particle energies do not correspond to physical
excitation energies, and in fact the widely used semi-local
xc-potentials significantly underestimate quasiparticle (QP) energy
gaps.\cite{Godby_PRB1988, new-developments} For molecules and
insulators better results can be obtained by replacing a fraction of
the local exchange potential with the non-local Hartree-Fock exchange
potential, as in the hybrid functionals. In the range-separated hybrids,
the non-local exchange is used only for the short-range part of the
potential. This improves the quality of semiconductor band structures
and leads to faster convergence with k-point sampling, albeit at the cost
of introducing an empirical cutoff radius. Still, the (range-separated)
hybrids tend to underestimate the role of exchange in systems with weak
screening, such as low-dimensional structures, and fail to account correctly
for the spatial variation in the screening at metal-insulator interfaces
(see below).

Many-body perturbation theory, on the other hand, offers a powerful and
rigorous framework for the calculation of quasiparticle (QP) excitations.
The key quantity is the electronic self-energy which is an energy-dependent
and spatially non-local analogous of the xc-potential of DFT. The self-energy
can be systematically approximated by summing certain classes of
perturbation terms to infinite order in the Coulomb interaction.  The
GW approximation \cite{Hedin_PR1965} is the simplest approximation of
this kind where the self-energy, $\Sigma$, is expanded to first order
in the screened interaction. Symbolically it takes the form
$\Sigma_{xc}=i G W$, where $G$ is the Green's function and
$W=\varepsilon^{-1}V$, is the screened interaction. Comparing the GW 
self-energy to the exchange potential, which can be
written as $V_x=i G V$, we see that the GW self-energy is
essentially a dynamically screened version of the exchange potential.

Apart from screening the static exchange
potential, the replacement of the bare Coulomb interaction by the
\emph{dynamically} screened potential introduces correlation effects
which accounts for the interaction of an electron (or a hole) with the
polarization charge that it induces in the medium.  This is a highly
nonlocal effect that becomes particularly evident at metal/insulator
interfaces such as a molecule on a metal surface or the
graphene/\textit{h}-BN interfaces studied in the present work.  For
these systems, the correlation takes the form of an image charge
effect that reduces the energy gap of the molecule or insulator by up to
several electron volts.\cite{Charlesworth_PRL1993, Inkson_JPhysC1973,
  Neaton_PRL2006, Garcia_PRB2009, Thygesen_PRL2009, Freysoldt_PRL2009}

The GW approximation has been applied with great success to a broad
class of materials ranging from bulk insulators, semiconductors and
metals to low-dimensional systems like nanoclusters, surfaces and
molecules (see e.g. the reviews of Refs. \onlinecite{qp-calculations,
  gw-method, DFTvsGW}).  Beyond the calculation of QP energies, the GW
method also serves as starting point for the calculation of optical
spectra from the Bethe-Salpeter equation (BSE)\cite{Albrecht_PRL1998,
  Salpeter-Bethe_PhysRev1951, Rohlfing-Louie_PRB2000, Yan_PRB2012} and
for quantitatively accurate modeling of electron transport at
metal-molecule interfaces where the alignment of the molecular energy
levels with the metal Fermi level is particularly important.
\cite{Thygesen_JChemPhys2007, Thygesen_PRB2008,
  Darancet_PRB2007, Strange_PRB2011, Strange_BJNano2011}

In principle, the GW self-energy should be evaluated
self-consistently.  However, due to the computational demands of such
an approach, non-selfconsistent (G$_0$W$_0$) calculations with the
initial $G_0$ obtained from the local density approximation (LDA) or
similar, have traditionally been preferred. Recently, fully
self-consistent GW calculations have been performed for molecular
systems yielding energies for the highest occupied orbitals with an absolute 
deviation from experiments of ~0.5 eV.\cite{Rostgaard_PRB2010,
  Blase_PRB2011} In comparison, the standard G$_0$W$_0$@LDA approach
was found to yield slightly lower accuracy, while better results were achieved
when starting from Hartree-Fock or hybrid calculations.\cite{Rostgaard_PRB2010, 
  Blase_PRB2011, Bruneval_JCTC2013, Caruso_PRB2012} For solids, earlier studies 
yielded contradictory  conclusions regarding the accuracy of self-consistent
versus non-selfconsistent GW calculations. More recently, the
quasiparticle selfconsistent GW method, in which the self-energy is evaluated
with a selfconsistently determined single-particle Hamiltonian,
has been shown to yield excellent results for solids.
\cite{Shishkin-Kresse_PRB2007, Faleev_PRL2004, Surh_PRB1991,
  Kotani_JPhysCondMat2007}

On the practical side, any implementation of the GW approximation has
to deal with similar numerical challenges. In addition to the already
mentioned G$_0$W$_0$ approximation, it is common practice to evaluate
the QP energies using first-order perturbation theory starting from the 
Kohn-Sham eigenvalues thereby
avoiding the calculation of off-diagonal matrix elements of the
self-energy. This approach is based on the assumption that the QP wave 
functions are similar to the Kohn-Sham wave functions.
As recently shown for a metal-molecule interface this is sometimes far from 
being the case.\cite{Strange_PRB2012}
Another common simplification is the use of a plasmon
pole approximation (PPA) for the dielectric function. The PPA leads to a 
considerable gain in efficiency by removing the need for
evaluating the dielectric function at all frequency points and allowing the 
frequency convolution of $G$ and $W$ in the GW self-energy to be carried out
analytically. In his original
paper, Hedin introduced a static COHSEX approximation to the full GW
self-energy. The COHSEX approximation is computationally efficient and clearly 
illustrates the 
physics described by the GW approximation. However, its
validity is limited to rather special cases and it should generally
not be used for quantitative calculations.

In this paper we document the implementation of the G$_0$W$_0$ method
in the GPAW open source electronic structure code.\cite{GPAW} GPAW is
based on the projector augmented wave method\cite{Bloechl_PRB1994,
  Bloechl_BMS2003} and supports both real space grid and plane wave
representation for high accuracy as well as numerical atomic orbitals
(LCAO) for high efficiency. The G$_0$W$_0$ implementation is based on
plane waves. The implementation supports both full frequency dependence (along
the real axis) as well as the plasmon-pole approximation of Godby and
Needs.\cite{Godby-Needs_PPA} For low dimensional systems, in particular
2D systems, a truncated Coulomb interaction should be  used to avoid the
long range interactions between periodically repeated unit cells.
For both solids, molecules and 2D systems, we find that the PPA gives
excellent results with significant reduction of the computational efforts.
In contrast, the static COHSEX and the PBE0 hybrid yield unsatisfactory
results. An interesting alternative to GW is offered by the local, orbital
dependent GLLBSC potential which explicitly adds the derivative
discontinuity to the Kohn-Sham energy gap.\cite{Gritsenko_PRA1995} The
GLLBSC band gaps for solids are found to lie on average within
$\unit[0.4]{eV}$ of the G$_0$W$_0$ values but give similar accuracy
when compared to experimental data.  The GLLBSC ionization potentials
of molecules are in average $\unit[1.5]{eV}$ below the G$_0$W$_0$
values.

The paper is organized as follows. Sec. \ref{sec:QPtheory} gives
a general introduction to the theory of quasiparticle states.
In Sec. \ref{sec:method}, we briefly review the central equations of
the G$_0$W$_0$ method in a plane wave basis and discuss some details
of our implementation. In Sec. \ref{sec:bulks}, we present results for
bulk semiconductors, insulators and metals, comparing with experiments
and previous calculations. The application to 2D systems is illustrated
in Sec. \ref{sec:2D} by the example of graphene on hexagonal boron nitride
and the importance of screening effects on the QP energies is discussed.
Finally, we test the implementation on finite systems by calculating
the ionization potential of a set of 32 small molecules in Sec.
\ref{sec:molecules}.

\section{Quasiparticle theory}\label{sec:QPtheory}

Quasiparticle states provide a rigorous generalization of the concept of 
single-particle orbitals to interacting electron systems. In this section we
provide a compact, self-contained introduction to the general theory of
quasiparticle states with a combined focus on physical interpretation and
mathematical rigor. This presentation is completely formal; in particular we
shall not discuss the physics and computation of specific self-energy
approximations. Our presentation is thus complementary to most other papers
on the GW method which tend to focus on the theory and derivation of the
GW self-energy within the framework of many-body Green's function theory. 
To avoid inessential mathematical complications, we shall make the assumption 
that the system under consideration is finite and the relevant excitations are
discrete.
 
\subsection{Definition of QP energies and wave functions}

We denote the $N$-particle many-body eigenstates and energies by 
$|\Psi^N_i\rangle$ and $E^N_i$, respectively. The occupied and unoccupied QP 
orbitals are denoted $|\psi^{\text{QP}}_{i-}\rangle$ and
$|\psi^{\text{QP}}_{i+}\rangle$, respectively.
These belong to the single-particle Hilbert space and are defined as:
\begin{eqnarray}
\label{eq.QP1}
\psi^{\text{QP}}_{i-}(\mathbf r)^* &=&
\langle \Psi_i^{N-1} | \hat \Psi(\mathbf r) | \Psi_0^N \rangle\\
\label{eq.QP2}
\psi^{\text{QP}}_{i+}(\mathbf r) &=&
\langle \Psi_i^{N+1} | \hat \Psi^{\dagger}(\mathbf r) | \Psi_0^N \rangle,
\end{eqnarray}
where $\hat \Psi(\mathbf r)$ and $\hat \Psi^{\dagger}(\mathbf r)$ are
the field operators annihilating and creating an electron at point $\mathbf r$,
respectively. The QP wave functions defined above are also sometimes referred
to as Lehman amplitudes or Dyson orbitals.
 
The corresponding QP energies are defined by
\begin{eqnarray}\label{eq.QPen1}
\varepsilon^{\text{QP}}_{i-}&=& E^N_0-E^{N- 1}_i\\ \label{eq.QPen2}
\varepsilon^{\text{QP}}_{i+}&=& E^{N+ 1}_i-E^N_0.
\end{eqnarray}
They represent the excitation energies of the $(N \pm 1)$-particle system
relative to $E^N_0$ and thus correspond to electron addition and removal
energies. It is clear that \mbox{$\varepsilon^{\text{QP}}_{i+}>\mu$} while
\mbox{$\varepsilon^{\text{QP}}_{i-}\leq \mu$} where
$\mu$ is the chemical potential. Having noted this, we can in fact drop the 
$+/-$ subscripts on the QP states and energies. We shall do that in most of the 
following to simplify the notation.
 
The fundamental energy gap is defined as
\begin{eqnarray}
E_{\text{gap}}&=&\varepsilon^{\text{QP}}_{0+}-\varepsilon^{\text{QP}}_{0-}\\
&=&E^{N+1}_0+E^{N-1}_0-2E^N_0.
\end{eqnarray}
We note that $E_{\text{gap}}$ can also be expressed within the framework of 
Kohn-Sham (KS) theory as
\begin{equation}
E_{\text{gap}}=\varepsilon^{\text{KS}}_{N+1}-\varepsilon^{\text{KS}}_{N}
+ \Delta_{xc},
\end{equation}
where $\varepsilon^{\text{KS}}_{n}$ are the (exact) Kohn-Sham energies and 
$\Delta_{xc}$ is the derivative discontinuity.\cite{Perdew_PRL1982}
 
 \subsection{Interpretation of QP wave functions}
 
Since the many-body eigenstates of an interacting electron system are not 
Slater determinants, the notion of single-particle orbitals is not well
defined \emph{a priori}. For weakly correlated systems we can, however,
expect that the single-particle picture applies to a good approximation.
To make this precise we ask to which extent the state $|\Psi^{N+1}_i\rangle$
can be regarded as a single-particle excitation from the groundstate, i.e.
to which extent it can be written on the form
$c^\dagger_\phi |\Psi^N_0\rangle$ when $\phi$ is chosen in an optimal way.
It turns out that the optimal $\phi$ is exactly the QP orbital.
This statement follows simply from the observation
\footnote{With the use of $c^\dagger_\phi = \int \! d\mathbf{r} \,
\phi^*(\mathbf r) \hat \Psi^{\dagger}(\mathbf r).$}
\begin{equation}\label{eq.QPint}
\langle \phi | \psi^{\text{QP}}_{i+} \rangle
= \langle \Psi_i^{N+1}|\hat c_\phi^\dagger | \Psi_0^N \rangle.
\end{equation}
Similarly, $|\psi^{-}_{i}\rangle$ is the orbital that makes
$\hat c_\phi|\Psi_0^N \rangle$ the best approximation to the excited state
$|\Psi_i^{N-1}\rangle$. Consequently, the QP wave function
$\psi^{\text{QP}}_{i\pm}$ is the single-particle orbital that best describes
the state of the "extra" electron/hole in the excited state
$|\Psi_i^{N\pm 1}\rangle$.
 
From Eq. (\ref{eq.QPint}) it follows that the norm of a QP orbital is a measure 
of how well the true excitation can be described as a single-particle 
excitation. Precisely,
\begin{equation}
\| \psi^{\text{QP}}_{i+} \| =
\max_\phi \Big \{ \langle \Psi_i^{N+1} | \hat  c_\phi^\dagger |
\Psi_0^N \rangle \quad , \quad \| \phi \| = 1 \rangle \Big \},
\end{equation}
and similarly for the norm of $\psi^{\text{QP}}_{i-}$.
 
The definition (\ref{eq.QP1}) implies a one-to-one correspondence between QP
states and the excited many-body states $|\Psi_i^{N\pm 1}\rangle $. Obviously,
most of the latter are not even approximately of the single-particle type.
These are characterized by a vanishing (or very small) norm of the
corresponding QP orbital. In case of non-interacting electrons the QP states
have norms 1 or 0. The former correspond to single excitations (Slater
determinants) of the form $c^\dagger _n|\Psi^N_0\rangle$ while the latter
correspond to multiple particle excitations, e.g.
$c^\dagger _n c^\dagger _m c^{\vphantom{\dagger}}_k| \Psi^N_0 \rangle$.
Strictly speaking the term "quasiparticle" should be used only for those 
$|\psi^{+/-}_{i}\rangle$ whose norm is close to 1. The number of such 
states and whether any exists at all, depends on the system. For weakly 
correlated systems, one can expect a one-to-one correspondence between
the QP states with norm $\sim 1$ and the single-particle states of some
effective non-interacting Hamiltonian, at least for the low-lying excitations.
 
\subsection{Quasiparticle equation and self-energy}

Below we show that QP states fulfill a generalized eigenvalue equation known as 
the QP equation, and we derive a useful expression for the norm of a QP state 
in terms of the self-energy.
 
The QP states and energies are linked to the single-particle Green's function
via the Lehmann spectral representation\cite{Bruus-Flensberg}
\begin{equation}\label{eq.spec1}
G(z) = \sum_i \frac{| \psi^{\text{QP}}_i \rangle \langle \psi^{\text{QP}}_i |}
{z - \varepsilon^{\text{QP}}_i},
\end{equation}
where $z$ is a complex number and it is understood that the sum runs over both 
occupied and unoccupied QP states. It follows that $G(z)$ is
analytic in the entire complex plane except for the real points 
$\varepsilon^{\text{QP}}_i$ which are simple poles. We note in passing that 
$G(z)$ equals the Fourier transform of the retarded (advanced) Green's function
in the upper (lower) complex half plane.
 
The Green's function also satisfies the Dyson equation
\begin{equation}
G(z)=[z-H_{0}-\Sigma_{xc}(z)]^{-1},
\end{equation}
where $H_0$ is the non-interacting part of the Hamiltonian including Hartree 
field and $\Sigma_{xc}$ is the exchange-correlation self-energy. The Dyson 
equation can be derived using many-body perturbation theory or it can simply be 
taken as the definition of the self-energy operator.
 
In the case where $\varepsilon^{\text{QP}}_i$ belongs to the discrete spectrum,
$\psi^{\text{QP}}_{i}$ and $\varepsilon^{\text{QP}}_{i}$ are solutions to the 
QP equation
\begin{equation}
\big [ H_0 + \Sigma_{xc}(\varepsilon^{\text{QP}}_i) \big ] |
\psi^{\text{QP}}_i \rangle = \varepsilon_i^{\text{QP}} |
\psi^{\text{QP}}_i \rangle.
\end{equation}
This follows from the residue theorem by integrating the equation 
\mbox{$[z - H_{0} - \Sigma_{xc}(z)] G(z) = 1$} along a complex contour enclosing
the simple pole $\varepsilon^{\text{QP}}_{i}$.
 
The operator $[H_0 + \Sigma_{xc}(z)]$ is non-Hermitian and is diagonalized by a 
set of non-orthogonal eigenvectors,
\begin{equation}
\big [ H_0 + \Sigma_{xc}(z)\big ] | \psi_n(z) \rangle
= \varepsilon_n(z) | \psi_n(z) \rangle.
\end{equation} 
Using these eigenvectors, the GF can be expressed in an alternative spectral 
form
\begin{equation}\label{eq.spec2}
G(z) = \sum_n \frac{| \psi_n(z) \rangle \langle \psi^n(z) |}
{z - \varepsilon_n(z)}.
\end{equation}
where $\{\psi^n(z)\}$ is the dual basis of $\{\psi_n(z)\}$ which by definition 
satisfies $\langle \psi_n(z)|\psi^m(z)\rangle = \delta_{nm}$.
\footnote{The dual basis functions are in fact the eigenvectors of the adjoint
operator $[H_0+\Sigma(z)]^\dagger$.} We shall take the functions $\psi_n(z)$ to
be normalized which also fixes the normalization of the dual basis.
 
In general, the vectors $\psi_n(z)$ do not have any physical meaning but are 
pure mathematical objects. An exception occurs for
$z = \varepsilon^{\text{QP}}_i$ where one of the vectors
$\psi_n(\varepsilon^{\text{QP}}_i)$ conincide with the QP orbital
$\psi^{\text{QP}_i}$ (except for normalization). We shall denote that vector by
$\psi_i(\varepsilon^{\text{QP}}_i)$, i.e.
\begin{equation}\label{eq.QP3}
| \psi_i(\varepsilon_i^{\text{QP}}) \rangle =
| \psi^{\text{QP}}_i \rangle / \| \psi_i^{\text{QP}} \|.
\end{equation}

By equating the matrix element $\langle \psi^i(z)|G(z)|\psi_i(z)\rangle$ 
evaluated using the two alternative spectral representations Eq. 
(\ref{eq.spec1}) and Eq. (\ref{eq.spec2}), and integrating along a contour 
enclosing the pole $\varepsilon_i^{\text{QP}}$, we obtain
\begin{equation}
\langle \psi^i(\varepsilon^{\text{QP}}_i) | \psi^{\text{QP}}_i \rangle
\langle \psi^{\text{QP}}_i | \psi_i(\varepsilon_i^{\text{QP}}) \rangle
= \frac{1}{1-\varepsilon_i'(\varepsilon^{\text{QP}}_i)},
\end{equation}
where the prime denotes the derivative with respect to $z$.
This result follows by application of the residue theorem. Using Eq. 
(\ref{eq.QP3}) it follows that the norm of the QP states is given by
\begin{eqnarray}
\| \psi^{\text{QP}}_i \|^2 &=&
\langle \psi_i(\varepsilon_i^{\text{QP}}) |
1 - \Sigma'_{xc}(\varepsilon_i^{\text{QP}}) |
\psi_i(\varepsilon_i^{\text{QP}}) \rangle ^{-1}\\
&\equiv &Z_i,
\end{eqnarray}
where we have used the Hellman-Feynman theorem to differentiate 
$\varepsilon_i(z) =
\langle \psi_i(z) | H_0 + \Sigma_{xc}(z) | \psi_i(z) \rangle$.
 
\subsection{Linearized QP equation}

Given a self-energy operator, one must solve the QP equation to obtain the QP
states and energies. This is complicated by the fact that the
self-energy must be evaluated at the QP energies which are not known
\emph{a priori}. Instead, one can start from an effective non-interacting 
Hamiltonian (in practice often the Kohn-Sham Hamiltonian),
\begin{equation}
[ H_0 + V_{xc} ] | \psi_i^s \rangle = \varepsilon_i^s |\psi_i^s \rangle,
\end{equation}
and treat $\Sigma_{xc}(z) - V_{xc}$ using first-order perturbation theory.
Thus we write $\varepsilon_i^{\text{QP}} = \varepsilon_i^s+\varepsilon_i^{(1)}$ 
with
\begin{eqnarray}
\varepsilon_i^{(1)} &=&
\langle \psi_i^s | \Sigma_{xc}(\varepsilon_i^{\text{QP}}) - V_{xc}
| \psi_i^s \rangle \\
&=&
\langle \psi_i^s | \Sigma_{xc}(\varepsilon_i^s) + (\varepsilon_i^{\text{QP}}
- \varepsilon_i^s) \Sigma'_{xc}(\varepsilon^s_i) - V_{xc} | \psi_i^s \rangle.
\end{eqnarray}
Rearranging this equation yields
\begin{equation}\label{eq.QPlin}
\varepsilon_i^{\text{QP}} = \varepsilon_i^s + Z_i^s \cdot
\langle \psi_i^s | \Sigma_{xc}(\varepsilon_i^s) - V_{xc} | \psi_i^s \rangle,
\end{equation}
where
\begin{equation}
Z_i^s = \langle \psi_i^s | 1 - \Sigma_{xc}'(\varepsilon_i^s) |
\psi_i^s\rangle ^{-1}
\end{equation}
approximates the true QP norm.
 
If $Z_i^s\ll 1$ we can conclude that $\psi_i^s$ is not a (proper) QP
state. There can be two reasons for this: (i) the electrons are
strongly correlated and as a consequence the QP picture does not
apply, or (ii) $\psi_i^s$ is not a good approximation to the true QP
wave function $\psi_i^{\text{QP}}$. While (i) is rooted in the physics
of the underlying electron system, reason (ii) merely says that the
Kohn-Sham orbital do not describe the true many-body exciations well. 
For an example where the QP picture is completely valid, i.e. all the QP
states have norms very close to 1 or 0, but where simple non-interacting
orbitals do not provide a good approximation to them, we refer to Ref.
\onlinecite{Strange_PRB2012}.

\section{G$_0$W$_0$ Approximation}\label{sec:method}

The self-energy of the GW approximation is given as a product of the Green's 
function and the screened Coulomb potential
and can be split into an exchange and a correlation part, $\Sigma_{\text{GW}} = 
V_x + \Sigma_c$, where $V_x$ is the non-local Hartree-Fock 
exchange potential.
The correlation contribution (which we from now on refer to as the self-energy 
$\Sigma = \Sigma_c$) is then evaluated by introducing 
the difference between the screened and the bare Coulomb potential 
$\overline{W} = W - V$:
\begin{equation}
  \Sigma(\mathbf{r} t, \mathbf{r}' t') = i G(\mathbf{r} t, \mathbf{r}' t')
  \overline{W}(\mathbf{r} t, \mathbf{r}' t'),
\label{eq:Sigma_rt}
\end{equation}
which becomes a convolution in frequency domain:
\begin{equation}
  \Sigma(\mathbf{r}, \mathbf{r}'; \omega) = \frac{i}{2 \pi} \int\!d\omega' \,
  G(\mathbf{r}, \mathbf{r}'; \omega + \omega') \overline{W}(\mathbf{r},
\mathbf{r}'; \omega').
\label{eq:Sigma_kn}
\end{equation}
In this way, the exchange and the correlation contributions can be treated 
separately at different levels of accuracy.
Additionally, the screened Coulomb potential approaches the bare one for large 
frequencies, so that $\overline{W}$ vanishes in this limit making the frequency 
integration numerically stable.

In the present G$_0$W$_0$ approach, the self-energy is constructed 
from Kohn-Sham wavefunctions $\left|n \mathbf{k}\right>$
and eigenvalues $\varepsilon_{n \mathbf{k}}^s$,
where $n$ and $\mathbf{k}$ denote band and k-point index, respectively.
Throughout this paper, spin indices are supressed in order to simplify the
notation.

Using the spectral representation for the Green's function in this
basis and Fourier transforming to reciprocal space,
the diagonal terms of the self-energy read:
\cite{Shishkin-Kresse_PRB2006}
\begin{eqnarray}
  \Sigma_{n \mathbf{k}} &\equiv& \left<n \mathbf{k} \middle|
  \Sigma(\omega) \middle|n \mathbf{k} \right>\nonumber\\
  &=& \frac{1}{\Omega} \sum\limits_{\mathbf{G} \mathbf{G}'}
  \sum\limits_{\vphantom{\mathbf{G}}\mathbf{q}}^{1. \text{BZ}}
  \sum\limits_{\vphantom{\mathbf{G}}m}^{\text{all}}
  \frac{i}{2 \pi} \int\limits_{-\infty}^\infty\!d\omega'\,
  \overline{W}_{\mathbf{G} \mathbf{G}'}(\mathbf{q}, \omega')\nonumber\\
  && \times \frac{\rho^{n \mathbf{k}}_{m \mathbf{k} - \mathbf{q}}(\mathbf{G})
  \rho^{n \mathbf{k}*}_{m \mathbf{k} - \mathbf{q}}(\mathbf{G}')}
  {\omega + \omega' - \varepsilon_{m \, \mathbf{k} - \mathbf{q}}^s
  + i \eta \, \text{sgn}(\varepsilon_{m \, \mathbf{k} - \mathbf{q}}^s - \mu)},
\label{eq:Sigma}
\end{eqnarray}
where $m$ runs over all bands, $\mathbf{q}$ covers the differences
between all k-points in the first Brillouin zone. The infinitesimal $\eta 
\rightarrow 0^+$
ensures the correct time-ordering of the Green's function, $\Omega =
\Omega_\text{cell} \cdot N_\mathbf{k}$ is the total crystal volume, and $\mu$ is
the chemical potential. The pair
density matrix elements are defined as:
\begin{equation}
  \rho^{n \mathbf{k}}_{m \mathbf{k} - \mathbf{q}}(\mathbf{G}) \equiv
  \left<n \mathbf{k} \middle| e^{i(\mathbf{q} + \mathbf{G})\mathbf{r}}
  \middle|m \, \mathbf{k} \!-\! \mathbf{q} \right>.
\label{eq:rho}
\end{equation}
The potential $\overline{W}_{\mathbf{G} \mathbf{G}'}(\mathbf{q},
\omega)$ is obtained from the symmetrized, time-ordered dielectric
function in the random phase approximation (RPA):
\begin{equation}
  \overline{W}_{\mathbf{G} \mathbf{G}'}(\mathbf{q}, \omega) =
  \frac{4 \pi}{|\mathbf{q} + \mathbf{G}|}
  \left( \epsilon^{-1}_{\mathbf{G} \mathbf{G}'}(\mathbf{q}, \omega)
  - \delta^{\vphantom{-1}}_{\mathbf{G} \mathbf{G}'} \right)
  \frac{1}{|\mathbf{q} + \mathbf{G}'|}.
\label{eq:W_GGqw}
\end{equation}
The calculation of the dielectric function in the GPAW code is
described in Ref. \onlinecite{Yan_PRB2011}.

The quasi-particle spectrum is then calculated with Eq. \ref{eq.QPlin}
using first-order perturbation theory in
$(\Sigma_{\text{GW}} - V_{xc})$, where $V_{xc}$
is the Kohn-Sham exchange-correlation potential:
\begin{equation}
  \varepsilon^{\text{QP}}_{n \mathbf{k}} = 
  \varepsilon_{n \mathbf{k}}^s + Z_{n \mathbf{k}}^s \cdot
  \text{Re} \left< n \mathbf{k}\middle| \Sigma(\varepsilon_{n \mathbf{k}}^s) +
  V_x - V_{xc} \middle| n \mathbf{k}\right>,
\label{eq:QP_kn}
\end{equation}
with a renormalization factor given by:
\begin{equation}
  Z_{n \mathbf{k}}^s = \left(1 - \text{Re}\left< n \mathbf{k}\middle|
  \Sigma'(\varepsilon_{n \mathbf{k}}^s)
  \middle| n \mathbf{k}\right>\right)^{-1},
\label{eq:Z_kn}
\end{equation}
where the derivative of the self-energy with respect to the
frequency is calculated analytically from
Eq. (\ref{eq:Sigma}). The calculation of the exact exchange 
potential within GPAW is described in Ref. \onlinecite{GPAW} using the plane 
wave expressions of Ref. \onlinecite{Sorouri_JChemPhys2006}.

As discussed in the previous section, this first-order approach,
i.e. using only the diagonal terms of the self-energy,
is based on the assumption that the true QP wave functions and energies
are similar to the Kohn-Sham wave functions and energies. To proceed beyond
this approximation one must evaluate also the off-diagonal terms of the
self-energy and invoke (partial) self-consistency. This is, however,
beyond the scope of the present work. Similarly, the effect of
electron-electron interactions on the QP lifetimes, which in principle
can be deduced from the imaginary part of the GW self-energy,
will not be considered in this study.

\subsection{Frequency grid}\label{subsec:frequency}

For a fully frequency-dependent GW calculation, the dielectric matrix
and thus the screened potential is evaluated on a user-defined grid of
real frequencies and the integration in Eq. (\ref{eq:Sigma}) is
performed numerically. The frequency grid is chosen to be linear up to
$\omega_{\text{lin}}$ with a spacing of $\Delta\omega$
which typically is set to $\unit[0.05]{eV}$.  Above $\omega_{\text{lin}}$ the
grid spacing grows linearly up to a maximum
frequency, $\omega_{\text{max}}$. In practice we set
$\omega_{\text{max}}$ to equal the maximum transition energy
and $\omega_{\text{lin}} \approx (1/4) \cdot \omega_{\text{max}}$ which
results in a few thousand frequency points.
Compared to a fully linear grid, the use of a non-uniform 
grid gives a computational speedup of around a factor $2 - 3$
without any loss of accuracy.  The broadening parameter $\eta$ is set
to $4 \Delta\omega$ to ensure a proper resolution of all spectral
features.

\subsection{Plasmon pole approximation}\label{subsec:ppa}

In the plasmon pole approximation (PPA), the frequency dependence of
the dielectric function $\epsilon^{-1}_{\mathbf{G}
  \mathbf{G}'}(\mathbf q,\omega)$ is modeled as a single pole
approximation:
\begin{align}
  \varepsilon^{-1}_{\mathbf{G}\mathbf{G}'}(\mathbf{q}, \omega) \, = \,
  & R _{\mathbf{G}\mathbf{G}'}(\mathbf{q})
  \left(\frac{1}{\omega - \tilde{\omega}_{\mathbf{G}\mathbf{G}'}(\mathbf{q})
  + i\eta} \right. \nonumber \\
  & \hphantom{R _{\mathbf{G}\mathbf{G}'}(\mathbf{q})} -
  \left. \frac{1}{\omega + \tilde{\omega}_{\mathbf{G}\mathbf{G}'}(\mathbf{q})
  - i\eta}\right).
\label{eq:PPA}
\end{align}
The plasmon frequency
$\tilde{\omega}_{\mathbf{G}\mathbf{G}'}(\mathbf{q})$ and the (real)
spectral function $R _{\mathbf{G}\mathbf{G}'}(\mathbf{q})$ are
determined by fitting this function to the dielectric matrix given at
the frequency points $\omega_1 = 0$ and $\omega_2 = i E_0$:
\begin{align}
  \tilde{\omega}_{\mathbf{G}\mathbf{G}'}(\mathbf{q}) = \, & E_0
  \sqrt{\frac{\varepsilon^{-1}_{\mathbf{G}\mathbf{G}'}(\mathbf{q}, \omega_2)}
  {\varepsilon^{-1}_{\mathbf{G}\mathbf{G}'}(\mathbf{q}, \omega_1) -
  \varepsilon^{-1}_{\mathbf{G}\mathbf{G}'}(\mathbf{q}, \omega_2)}},\\
  R _{\mathbf{G}\mathbf{G}'}(\mathbf{q}) = \, &
  -\frac{\tilde{\omega}_{\mathbf{G}\mathbf{G}'}(\mathbf{q})}{2}
  \varepsilon^{-1}_{\mathbf{G}\mathbf{G}'}(\mathbf{q}, \omega_1).
\label{eq:PPA_wR}
\end{align}
Using the relation
\begin{equation}
  \lim_{\eta \rightarrow 0^+} \frac{1}{x \pm i\eta} =
  \mathcal{P}\left\{\frac{1}{x}\right\} \mp i\pi\delta(x),
\label{eq:Cauchy}
\end{equation}
where $\mathcal{P}$ denotes the Cauchy principal value, the spectral
function of the screened potential,
$\text{Im}\left\{\overline{W}_{\mathbf{G} \mathbf{G}'}(\mathbf
  q,\omega)\right\}$, is simply a delta function at the plasmon
frequencies $\pm \tilde \omega_{\mathbf{G} \mathbf{G}'}(\mathbf q)$.
Similarily, the relation (\ref{eq:Cauchy}) can be used in Eq. (\ref{eq:Sigma}) 
allowing the GW self-energy to be evaluated analytically.

The PPA is expected to be a good approximation, when the overall structure of 
the dielectric function is dominated by a single (complex) pole. The true 
dielectric function will show variations on a finer scale. However, these are 
averaged out by the frequency integration in Eq. (\ref{eq:Sigma}).
In practice, we set the free parameter, $E_0$, to $\unit[1]{\text{Hartree}}$ in 
all our calculations and we find results to be insensitive to variations
of around $\unit[0.5]{\text{Hartree}}$.

\subsection{Static COHSEX}\label{subsec:COHSEX}

By setting $\omega - \varepsilon_{m \, \mathbf{k}  - \mathbf{q}} = 0$
in Eq. (\ref{eq:Sigma}), the self-energy becomes
frequency-independent and can be split into two parts, named Coulomb
hole and Screened exchange.\cite{Hybertsen-Louie_PRB86} The first term
arises from the poles of the screened potential and describes the
local interaction of an electron with its induced charge:
\begin{equation}
  \Sigma^{\text{COH}} = \frac{1}{2} \delta(\mathbf{r} - \mathbf{r}')
  \left(W(\mathbf{r}, \mathbf{r}'; \omega=0)
  - V(\mathbf{r}, \mathbf{r}')\right).
\end{equation}
The plane wave expression for a matrix element on a Bloch state $|n\mathbf 
k\rangle$ becomes
\begin{equation}
  \Sigma_{n \mathbf{k}}^{\text{COH}} =
  \frac{1}{2 \Omega} \sum\limits_{\mathbf{G} \mathbf{G}'}
  \sum\limits_{\vphantom{\mathbf{G}}\mathbf{q}}
  \sum\limits_{\vphantom{\mathbf{G}}m}^{\text{all}}
  \overline{W}_{\mathbf{G} \mathbf{G}'}^{\vphantom{-1}}(\mathbf{q}, 0)
  \rho^{n \mathbf{k}}_{m \mathbf{k} - \mathbf{q}}(\mathbf{G})
  \rho^{n \mathbf{k}*}_{m \mathbf{k} - \mathbf{q}}(\mathbf{G}').
\label{eq:SigmaCOH}
\end{equation}
The second term originates from the poles of the Green's function and
is identical to the exchange term in Hartree-Fock theory with the
Coulomb kernel replaced by the screened interaction:
\begin{equation}
  \Sigma^{\text{SEX}} = - \sum\limits_j^{\text{occ}} \phi_j^*(\mathbf{r})
  \phi_j^{\vphantom{*}}(\mathbf{r}') W(\mathbf{r}, \mathbf{r}'; \omega=0),
\end{equation}
which yields the matrix element
\begin{equation}
  \Sigma_{n \mathbf{k}}^{\text{SEX}} =
  - \frac{1}{\Omega} \sum\limits_{\mathbf{G} \mathbf{G}'}
  \sum\limits_{\vphantom{\mathbf{G}}\mathbf{q}}
  \sum\limits_{\vphantom{\mathbf{G}}m}^{\text{occ}}
  W_{\mathbf{G} \mathbf{G}'}^{\vphantom{-1}}(\mathbf{q}, 0)
  \rho^{n \mathbf{k}}_{m \mathbf{k} - \mathbf{q}}(\mathbf{G})
  \rho^{n \mathbf{k}*}_{m \mathbf{k} - \mathbf{q}}(\mathbf{G}').
\label{eq:SigmaSEX}
\end{equation}
The quasi-particle energies are then given as
\begin{equation}
  \varepsilon^{\text{QP}}_{n \mathbf{k}} = 
  \varepsilon_{n \mathbf{k}}^s + \left< n \mathbf{k}\middle|
  \Sigma^{\text{SEX}} + \Sigma^{\text{COH}} - V_{xc}
  \middle|n \mathbf{k}\right>.
\label{eq:QPCOHSEX}
\end{equation}

\subsection{Coulomb divergence}\label{subsec:divergence}

For $\mathbf{q} \rightarrow 0$, the head, $\overline{W}_{\mathbf{0}
  \mathbf{0}}(\mathbf q)$, and wings, $\overline{W}_{\mathbf{G}
  \mathbf{0}}(\mathbf q), \overline{W}_{\mathbf{0} \mathbf{G}'}(\mathbf q)$, of
the screened potential diverge as $1/q^2$ and $1/q$, respectively. These 
divergences are, however, integrable. In the limit of 
a very fine k-point sampling we have $\sum_{\mathbf{q}} \rightarrow
\frac{\Omega}{(2\pi)^3} \int\! d q \, 4 \pi q^2$, and thus we can replace the
$\mathbf{q} = 0$ term in the $q$-sum of Eq. (\ref{eq:Sigma}) by an integral 
over a sphere in
reciprocal space with volume $\Omega_{\text{BZ}} /
N_{\mathbf{k}}$. The head and wings of the screened potential then
take the form
\begin{align}
  \overline{W}_{\mathbf{00}}(\mathbf{q}=0, \omega) = \, & \frac{2\Omega}{\pi}
  \left(\frac{6\pi^2}{\Omega}\right)^{1/3}
  \left[\varepsilon^{-1}_{\mathbf{00}}(\mathbf{q} \rightarrow 0,
    \omega) - 1\right],
  \label{eq:Wdiv_head} \\
  \overline{W}_{\mathbf{G0}}(\mathbf{q}=0, \omega) = \, &
  \frac{1}{|\mathbf{G}|}\frac{\Omega}{\pi}
  \left(\frac{6\pi^2}{\Omega}\right)^{2/3}
  \varepsilon^{-1}_{\mathbf{G0}}(\mathbf{q} \rightarrow 0, \omega),
\label{eq:Wdiv_wings}
\end{align}
with the dielectric function evaluated in the optical limit.\cite{Yan_PRB2011}

\subsection{Coulomb truncation}\label{subsec:truncation}

In order to avoid artificial image effects in supercell calculations
of systems which are non-periodic in one direction (2D systems), we
follow Ref. \onlinecite{Rozzi_PRB2006} and cut off the Coulomb
interaction by a step function in the non-periodic direction
($z$-axis)
\begin{equation}
  \tilde{v}^{2D}(\mathbf{r}) = \frac{\theta(R - |r_z|)}{|\mathbf{r}|},
\label{eq:vcut_r}
\end{equation}
where $R$ is the truncation length. In reciprocal space, this becomes
\begin{equation}
  \tilde{v}^{2D}(\mathbf{G}) = \frac{4 \pi}{\mathbf{G}^2}
  \left[1 + e^{-G_{\|} R} \left(\frac{G_z}{G_{\|}} \sin(G_z R)
  - \cos(|G_z| R)\right) \right],
\label{eq:vcut_G}
\end{equation}
where $G_{\|}$ and $G_z$ are the parallel and perpendicular components
of $\mathbf{G}$, respectively.  By setting $R$ to half the length of
the unit cell in $z$-direction, this simplifies
to\cite{Ismail-Beigi_PRB2006}
\begin{equation}
  \tilde{v}^{2D}(\mathbf{G}) = \frac{4 \pi}{\mathbf{G}^2}
  \left( 1 - e^{-G_{\|} R} \cos(|G_z| R) \right).
\label{eq:vcut_Gz}
\end{equation}
Since Eqs. (\ref{eq:vcut_G}) and (\ref{eq:vcut_Gz}) are not well
defined for $G_{\|} \rightarrow 0$, we have to evaluate these terms by
numerical integration:
\begin{equation}
  \tilde{v}^{\text{2D}}(G_{\|}=0) = \frac{1}{\Omega'} \int\limits_{\Omega'} \!
  d \mathbf{q}' \, \tilde{v}^{\text{2D}}(G_z + \mathbf{q}'),
\label{eq:vcut_num}
\end{equation}
where $\Omega'$ is a small BZ volume around $G_{\|} = 0$. This
integral is well-defined and converges easily for a fine grid
$\mathbf{q}'$ not containing the $\Gamma$-point.

We mention that other methods have been applied to correct for the
spurious long rage interaction in GW calculations for
surfaces.\cite{Freysoldt_CPC2007, Freysoldt_PRB2008}

\subsection{Computational details}\label{subsec:computational}

The calculation of one matrix element of the self-energy of Eq.
(\ref{eq:Sigma}) scales as $N_\omega^{\vphantom{2}} \cdot
N_b^{\vphantom{2}} \cdot N_k^2 \cdot N_G^2$ with number of frequency
points, bands, k-points and plane waves, respectively.  The code is
parallelized over $\mathbf{q}$ vectors. For calculations
including the $\Gamma$-point only, i. e. isolated systems, full
parallelization over bands is used instead. Therefore, the
computational time scales linearly with the number of cores.  The
screened potential $\overline{W}_{\mathbf{G} \mathbf{G}'}(\mathbf{q},
\omega)$ is evaluated separately for every $\mathbf{q}$ as an array in
$\mathbf{G}$, $\mathbf{G}'$ and $\omega$. For large numbers of plane
waves and frequency points, this array can be distributed onto
different cores, thus reducing the memory requirement on every core.

In practice, the use of the plasmon pole approximation gives a
computational speedup of a factor of 5 - 20 on average compared to a
full frequency calculation. For both methods (PPA and full frequency
integration), the computational time spent on the evaluation of the
dielectric matrix and on the calculation of the quasi-particle
spectrum from the screened potential is comparable.

\section{Solids}\label{sec:bulks}

As a first application, we calulate the band structures of ten simple
semiconductors and insulators ranging from \sffamily Si \rmfamily to
\sffamily LiF \rmfamily thus covering a broad range of band gap sizes
of both direct and indirect nature.  We compare the different
approximation schemes within non-selfconsistent GW, namely (i) full
frequency dependence (ii) plasmon pole approximation and (iii) static
COHSEX. In all these cases the self-energy is calculated with orbitals
and single-particle energies obtained from an LDA calculation, i.e.
G$_0$W$_0$@LDA. In addition we perform non-selfconsistent Hartree-Fock
(HF), as well as PBE0 hybrid calculations in both cases using LDA
orbitals. Finally, we compare to self-consistent
GLLBSC\cite{Gritsenko_PRA1995, Kuisma_PRB2010} calculations.  The
GLLBSC is based on the PBEsol correlation potential and uses an
efficient approximation to the exact exchange optimized effective
potential which allows for explicit evaluation of the derivative
discontinuity, $\Delta_{\text{xc}}$. We have recently applied the
GLLBSC in computational screening studies of materials for
photo-catalytic water splitting.\cite{Castelli_EES2012a,
  Castelli_EES2012b} Here we present a systematic assessment of its
performance by comparing to experiments and GW results for various
types of systems.

\begin{table}[t]
\begin{centering}
\caption{\label{tab:bulkstructure}Geometric stuctures.}
\begin{tabularx}{0.85\columnwidth}{l @{\hspace{1cm}} l @{\hspace{1cm}} c}
\hline\hline
			 & structure  & lattice constant in \AA \bigstrut\\
\hline
\sffamily Si   \rmfamily & diamond    & 5.431 \\
\sffamily InP  \rmfamily & zincblende & 5.869 \\
\sffamily GaAs \rmfamily & zincblende & 5.650 \\
\sffamily AlP  \rmfamily & zincblende & 5.451 \\
\sffamily ZnO  \rmfamily & zincblende & 4.580 \\
\sffamily ZnS  \rmfamily & zincblende & 5.420 \\
\sffamily C    \rmfamily & diamond    & 3.567 \\
\sffamily BN   \rmfamily & zincblende & 3.615 \\
\sffamily MgO  \rmfamily & rocksalt   & 4.212 \\
\sffamily LiF  \rmfamily & rocksalt   & 4.024 \\
\hline\hline
\end{tabularx}
\end{centering}
\end{table}
The bulk structures and the used lattice constants are listed in table
\ref{tab:bulkstructure}.

\begin{figure}[t]
\begin{centering}
\includegraphics[width=0.85\columnwidth,clip=]{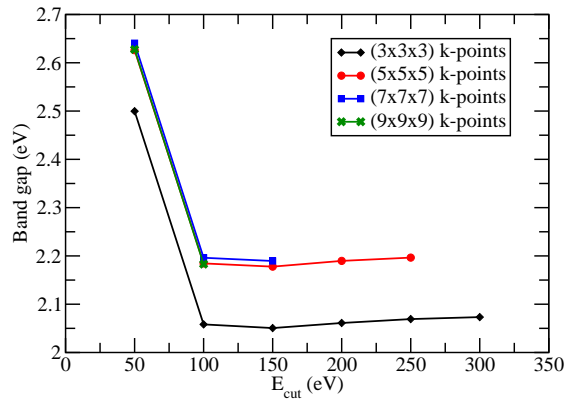}
\par\end{centering}
\caption{\label{fig:ZnOgap}(Color online). Convergence of the band gap of zinc
oxide for G$_0$W$_0$@LDA with the plasmon pole approximation.
The number of bands is chosen equally to the number of plane waves
corresponding to the respective cutoff energy, i. e. $\unit[300]{eV}$ equal
$\sim 1100$ plane waves and bands.}
\end{figure}
All calculations were performed with the GPAW code which is based on
the projector augmented wave method and supports both real space and
plane wave representations.  In the present work only the plane wave
basis set has been used. The same set of parameters is used for
the calculation of the dielectric matrix and the self-energy.  For all
GW calculations, convergence with respect to the plane wave cutoff,
number of unoccupied bands and k-points has been tested carefully, together 
with the size of the frequency grid for the full frequency calculations.  As an 
example, Fig. \ref{fig:ZnOgap} shows the dependence of the
G$_0$W$_0$ band gap of zinc oxide on the plane wave
cutoff and the number of k-points.  For cutoff energies above
$\unit[100]{eV}$ (corresponding to around 200 plane waves and bands),
the value of the band gap is converged to within $\unit[0.02]{eV}$, whereas 
increasing the number of k-points results in a
constant shift.  For all the solids we have investigated, the band gap
is well converged with $E_{\text{cut}} = \unit[200]{} -
\unit[300]{eV}$ and a few hundred empty bands.  For materials with
direct band gaps ($9\times 9\times 9$) k-points was found to be sufficient, 
whereas
for \sffamily AlP\rmfamily, \sffamily BN\rmfamily, \sffamily C
\rmfamily, \sffamily Si \rmfamily and \sffamily ZnS\rmfamily, which
have indirect gaps, ($15\times 15\times 15$) k-points were used in order to 
clearly resolve the conduction band minimum.

\begin{figure}[t]
\begin{centering}
\includegraphics[width=0.85\columnwidth,clip=]{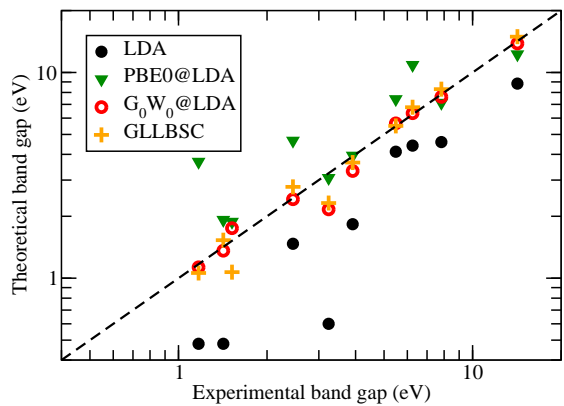}
\par\end{centering}
\caption{\label{fig:bandgaps}(Color online). Comparison of calculated
  and experimental band gaps for the solids listed in
  Tab. \ref{tab:bulkstructure}.  The numerical values are listed in
  Tab. \ref{tab:bandgaps}. A logarithmic scale is used for better
  visualization.  'G$_0$W$_0$@LDA' refers to the fully
  frequency-dependent non-selfconsistent GW based on LDA.  The
  PBE0 results are obtained non-selfconsistently using LDA orbitals.}
\end{figure}

The results for the band gaps are summarized in Fig. \ref{fig:bandgaps} and 
Table \ref{tab:bandgaps} along with experimental
data.  The last row shows the mean absolute errors (MAE) of each
method relative to experiment.

\begin{table*}[t]
  \caption{\label{tab:bandgaps} Band gaps in eV. The type of gap is indicated in
the last column. The last row gives the mean absolute error compared to
experiment.
    Experimental data is taken from Ref. \onlinecite{Fuchs_PRB2007}. Note that
the experimental data for \sffamily ZnO \rmfamily refers to the wurtzite
structure.
    We find the calculated band gap to be around $\unit[0.1]{eV}$ smaller in the
zincblende than in the wurtzite structure for both LDA, G$_0$W$_0$ and GLLBSC.
    Experimental gap for \sffamily InP \rmfamily taken from Ref.
\onlinecite{Vurgaftman_JApplPhys2001}.}
\begin{centering}
\begin{tabularx}{\textwidth}{X r@{.}l r@{.}l r@{.}l r@{.}l r@{.}l r@{.}l r@{.}l
r@{.}l r}
\toprule
\hline\hline
&\multicolumn{2}{c}{}   &\multicolumn{2}{c}{}      &\multicolumn{2}{c}{}       
&\multicolumn{6}{c}{G$_0$W$_0$@LDA} &\multicolumn{2}{c}{}
&\multicolumn{2}{c}{}\bigstrut\\
\cline{8-13}
&\multicolumn{2}{c}{LDA}&\multicolumn{2}{c}{HF@LDA}&\multicolumn{2}{c}{PBE0@LDA}
&\multicolumn{2}{c}{COHSEX}
&\multicolumn{2}{c}{PPA}&\multicolumn{2}{c}{dyn}   &\multicolumn{2}{c}{GLLBSC} 
&\multicolumn{2}{c}{experiment} & \bigstrut\\
\hline
\sffamily Si   \rmfamily &  0&48 &  5&26 	 &  3&68         &  0&56        
&  1&09 &  1&13 &  1&06 &  1&17 & indirect \\
\sffamily InP  \rmfamily &  0&48 &  5&51 	 &  1&92         &  1&99$^{(a)}$
&  1&38 &  1&36 &  1&53 &  1&42 & direct   \\
\sffamily GaAs \rmfamily &  0&38 &  5&46 	 &  1&88$^{(b)}$ &  3&77$^{(c)}$
&  1&76 &  1&75 &  1&07 &  1&52 & direct   \\
\sffamily AlP  \rmfamily &  1&47 &  7&15 	 &  4&66         &  1&88        
&  2&38 &  2&42 &  2&78 &  2&45 & indirect \\
\sffamily ZnO  \rmfamily &  0&60 & 10&42$^{(d)}$ &  3&07$^{(e)}$ &  0&10        
&  2&20 &  2&24 &  2&32 &  3&44 & direct   \\
\sffamily ZnS  \rmfamily &  1&83 &  9&43 	 &  3&94$^{(f)}$ &  1&52        
&  3&28 &  3&32 &  3&65 &  3&91 & direct   \\
\sffamily C    \rmfamily
&\hphantom{0000}4&12\hphantom{000}&\hphantom{000}11&83\hphantom{000}&\hphantom{
0000}7&42\hphantom{000}&\hphantom{0000}6&51\hphantom{000}
			
&\hphantom{0000}5&59\hphantom{000}&\hphantom{0000}5&66\hphantom{000}&\hphantom{
0000}5&50\hphantom{000}&\hphantom{0000}5&48\hphantom{000} & indirect \\
\sffamily BN   \rmfamily &  4&41 & 13&27 	 & 10&88         &  7&08        
&  6&30 &  6&34 &  6&78 &  6&25 & indirect \\
\sffamily MgO  \rmfamily &  4&59 & 14&84 	 &  7&12         & 10&30        
&  7&44 &  7&61 &  8&30 &  7&83 & direct   \\
\sffamily LiF  \rmfamily &  8&83 & 21&86 	 & 12&25         & 16&02        
& 13&64 & 13&84 & 14&93 & 14&20 & direct   \\
\hline
MAE			 &  2&05 &  5&74 	 &  1&52         &  1&59	
&  0&35 &  0&31 &  0&41 &   &   \bigstrut\\
\hline\hline
\bottomrule
\end{tabularx}
\end{centering}
\flushleft
$^{(a)}$COHSEX predicts an indirect band gap of $\unit[1.73]{eV}$.\\
$^{(b)}$PBE0 predicts an indirect band gap of $\unit[1.79]{eV}$.\\
$^{(c)}$COHSEX predicts an indirect band gap of $\unit[1.07]{eV}$.\\
$^{(d)}$HF predicts an indirect band gap of $\unit[9.73]{eV}$.\\
$^{(e)}$PBE0 predicts an indirect band gap of $\unit[2.83]{eV}$.\\
$^{(f)}$PBE0 predicts an indirect band gap of $\unit[3.80]{eV}$.
\end{table*}

As expected LDA predicts much too small band gaps with relative errors
as large as 400 \% in the case of \sffamily GaAs\rmfamily.  In
contrast HF greatly overestimates the band gap for all systems
yielding even larger relative errors than LDA and with absolute errors
exceeding 7 eV. The failure of HF is particularly severe for systems
with narrow band gaps like \sffamily Si \rmfamily and \sffamily InP
\rmfamily where the relative error is up to \mbox{500\%} whereas the error for
the large gap insulator \sffamily LiF \rmfamily is \mbox{50\%}.  This
difference can be understood from the relative importance of screening
(completely neglected in HF) in the two types of systems.

The PBE0 results lie in between LDA and HF with band gaps lying
somewhat closer to the experimental values, however, still
significantly overestimating the size of the gap for systems with
small to intermediate band gap.

The inclusion of static screening within the COHSEX approximation
significantly improves the bare HF results.  However, with a MAE of
$\unit[1.59]{eV}$, the results are still unsatisfactory and there
seems to be no systematic trend in the deviations from experiments,
except for a slightly better performance for materials with larger
band gaps. We mention that a detailed discussion of the drawbacks of
COHSEX and how to correct its main deficiencies can be found in Ref.
\onlinecite{Kang-Hybertsen_PRB2010}.  In Ref.
\onlinecite{Bruneval_PRB06}, the static COHSEX approximation was
explored as a starting point for G$_0$W$_0$ calculations and compared
to quasi-particle self-consistent GW calculations. However, no
systematic improvement over the LDA starting point was found.

Introducing dynamical screening in the self-energy brings the band
gaps much closer to the experimental values. The G$_0$W$_0$
calculations with the PPA and full frequency dependence yield almost
identical results, with only small deviations of about
$\unit[0.2]{eV}$ for the large band gap systems \sffamily LiF
\rmfamily and \sffamily MgO\rmfamily, where the fully
frequency-dependent method performs slightly better.

Our results agree well with previous works for G$_0$W$_0$ calculations
using LDA\cite{Kotani_SSC2002} and PBE\cite{Shishkin-Kresse_PRB2007}
as starting points with mean absolute errors of $\unit[0.31]{}$ and
$\unit[0.21]{eV}$ in comparison, respectively.  Compared to
Ref. \onlinecite{Shishkin-Kresse_PRB2007}, the only significant
deviations can be seen for \sffamily GaAs \rmfamily and the wide gap
systems, where our calculated band gaps are somewhat larger. We expect
that this is due to the difference between LDA and PBE as starting
point.  The values reported in Ref. \onlinecite{Kotani_SSC2002} are
all smaller than ours.  A more detailed comparison is, however,
complicated because of the differences in the implementations:
Ref. \onlinecite{Kotani_SSC2002} uses a mixed basis set in an
all-electron LMTO framework.
We note that for \sffamily LiF\rmfamily, the calculated 
band gap is strongly dependent on the lattice constant. With only a
slightly smaller lattice constant of $\unit[3.972]{\text{\AA}}$,
which is the experimental value corrected for zero-point anharmonic
expansion effects,\cite{Harl_PRB2010} the quasiparticle gap
increases by $\unit[0.4]{eV}$.

One well-known problematic case for the GW approximation is \sffamily ZnO 
\rmfamily (both in the zincblende and the wurtzite structure). The
calculated band gap in the present study at the G$_0$W$_0$@LDA level is about 
$\unit[1]{eV}$ too low which is consistent with other
previous G$_0$W$_0$ studies.\cite{Usuda_PRB2002, Dixit_JPhysCondMat2010, 
Rinke_NewJPhys2005, Fuchs_PRB2007} Recent
G$_0$W$_0$ calculations employing pseudopotentials and the 
PPA\cite{Shih_PRL2010} as well as all-electron
G$_0$W$_0$\cite{Friedrich_PRB2011} have attributed this discrepancy to
a very slow convergence of the band gap with respect to the number of bands.  
This is, however, not in agreement with our PAW based
calculations which are well converged with a cutoff energy of $\unit[100]{eV}$ 
and around 200 bands. We note that semi-core \textit{d}-states of zinc are
explicitly included in our calculations.
The large differences of the results and the convergence behaviour compared to 
Ref. \onlinecite{Shih_PRL2010} are most likely due to the use of different
models for the plasmon pole approximation. As discussed in Ref.
\onlinecite{Stankovski_PRB2011}, the use of a model dielectric function
which fulfills Johnson's \textit{f}-sum rule (as the PPA of Hybertsen and 
Louie)\cite{Hybertsen-Louie_PRB86}
leads to a very slow convergence of the band gap of \sffamily ZnO \rmfamily 
with respect to the number of plane waves and unoccupied bands
and gives a result which is $\unit[1]{eV}$ higher than obtained with the fully 
frequency dependent method.
With the PPA of Godby and Needs on the other hand, results converge 
considerably faster and agree remarkably well with the frequency dependent 
method.

Our results are consistent with Ref. \onlinecite{Shishkin-Kresse_PRB2007} who 
attributed the underestimation of the gap to the starting point
(PBE in their case) and also showed that the QP-sc GW method yields
a band gap of $\unit[3.20]{eV}$  in very good agreement with experiment.

The band gaps denoted GLLBSC in Table \ref{tab:bandgaps} have been
obtained as the self-consistently determined Kohn-Sham band gap of a
GLLBSC calculation with the estimated derivative discontinuity
$\Delta_{xc}$ added. Compared to G$_0$W$_0$, this approach
yields a slightly lower accuracy compared to experiment. On the other
hand, the much lower computational cost of the GLLBSC (which is
comparabe to LDA) makes this method very attractive for band structure
calculations of large systems.

We conclude that even single-shot GW calculations with the plasmon
pole approximation reproduce the experimental results to $~
\unit[0.2]{eV}$ for most of the semiconductors.  The largest
deviations are observed for \sffamily ZnO \rmfamily and \sffamily LiF
\rmfamily where the computed band gaps are around $\unit[1]{}$ and
$\unit[0.5]{eV}$ too small, respectively.  Both of these systems have
strong ionic character and LDA is presumably not a good starting point
-- in particular the LDA wave functions might be too delocalized.  In
such cases, a different starting point based on e.g. a hybrid or LDA+U
might yield better results although a systematic improvement seems
difficult to achieve in this way.\cite{Shishkin-Kresse_PRB2007}

\begin{figure}[t]
\begin{centering}
\includegraphics[width=0.85\columnwidth,clip=]{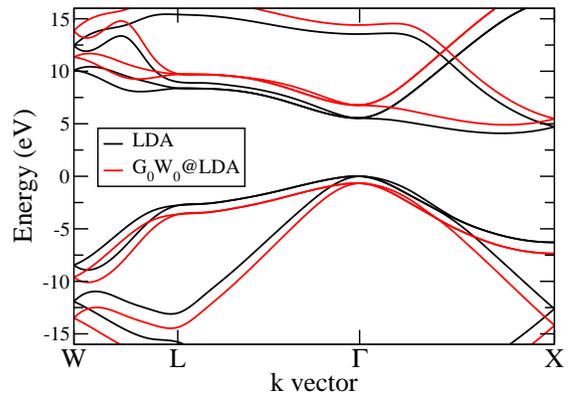}
\par\end{centering}
\caption{\label{fig:C_bandstructure}(Color online). Band structure of
  diamond calculated with LDA (black) and G$_0$W$_0$ (red). The bands
  have been interpolated by splines from a (15x15x15) k-point
  sampling.  The band gap is indirect between the $\Gamma$ point and
  close to the X point with a value of $\unit[4.12]{eV}$ and
  $\unit[5.66]{eV}$ for LDA and G$_0$W$_0$, respectively.}
\end{figure}

In Fig. \ref{fig:C_bandstructure}, we compare the band structure of
diamond obtained with the LDA and G$_0$W$_0$@LDA approximation.  The
valence band maximum occurs at the $\Gamma$-point and the conduction
band minimum is situated along the $\Gamma$--X-direction, resulting in
an indirect band gap of $4.1$ and $\unit[5.7]{eV}$, respectively.  We
can see that the main effect of the G$_0$W$_0$ approximation lies in
an almost constant shift of the LDA bands: Occupied bands are moved to
lower energies, whereas the unoccupied bands are shifted up.  This is
thus an example where the effect of G$_0$W$_0$ is well described by a
simple scissors operator.

\begin{figure}[t]
\begin{centering}
\includegraphics[width=0.85\columnwidth,clip=]{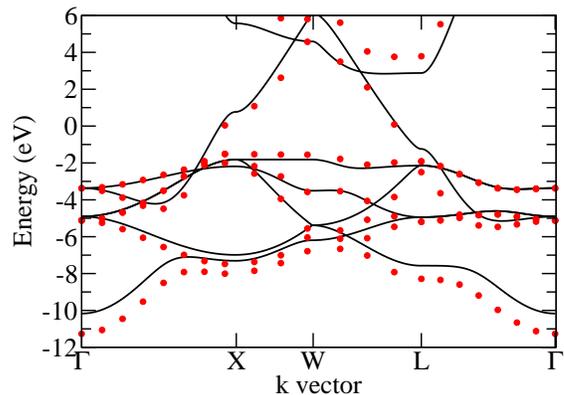}
\par\end{centering}
\caption{\label{fig:Au_bandstructure}Band structure of
  fcc gold calculated with LDA (black lines) and G$_0$W$_0$@LDA with PPA (red 
dots).
  (45x45x45) and (15x15x15) k-points have been used for LDA and GW, 
respectively.
  The bands are aligned to the respective Fermi level.}
\end{figure}

Finally, we present the calculated band structure of gold in Fig. 
\ref{fig:Au_bandstructure} as one example for a metallic system. The lattice 
parameter used for the fcc
structure is $\unit[4.079]{\text{\AA}}$. The effect of GW is a small broadening 
of the occupied \textit{d}-bands, with the top being shifted slightly 
up and the bottom down in energy.
The change in the low-lying \textit{s}-band and the unoccupied \textit{s-p} 
band are significantly larger and inhomogeneous.
Our band structure agrees well with the calculations of Ref. 
\onlinecite{Rangel_PRB2012} with use of the plasmon pole approximation
and exclusion of 5\textit{s} and 5\textit{p} semicore states.
In Ref. \onlinecite{Rangel_PRB2012} it was also shown that QP 
self-consistent GW approximation shifts the $d$-band down by $\unit[0.4]{eV}$
relative to PBE in good agreement with experiments.

\section{2D structures}\label{sec:2D}

\begin{figure}[t]
\begin{centering}
\includegraphics[width=0.85\columnwidth,clip=]{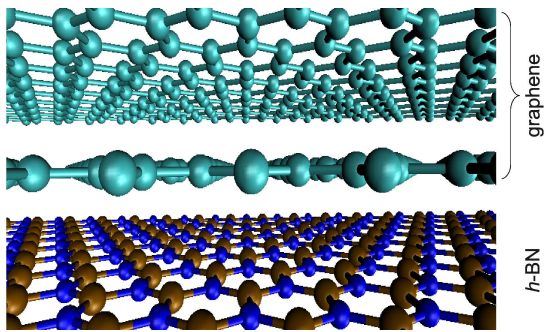}
\par\end{centering}
\caption{\label{fig:graphene-BN_scheme}Schematic picture
  of graphene/\textit{h}-BN interface.}
\end{figure}

In this section we investigate the quasiparticle band structure of a
two-dimensional structure composed of a single layer of
hexagonal-boron nitride (\textit{h}-BN) adsorbed on $N$ layers of
graphene (as sketched in Fig. \ref{fig:graphene-BN_scheme} for $N=2$).
Such 2D heterostructures have recently attracted much
attention due to their unique physical properties and potential
application in the next-generation electronic and photonic devices.
\cite{Ponomarenko_NaturePhys2011, Wang_ElectronDeviceLett2011,
  Haighl_NatureMat2012, Britnell_Science2012}

Since graphene and \textit{h}-BN are hexagonal structures with almost
the same lattice constant, \textit{h}-BN serves as a perfect substrate
for graphene.\cite{Decker_NL2011} Based on LDA total energy
calculations we find the most stable structure to be the configuration
with one carbon over the \sffamily B \rmfamily atom and the other
carbon centered above a \textit{h}-BN hexagon (equivalent to
configuration (c) of Ref. \onlinecite{Giovannetti_PRB07}) with a layer
separation of $\unit[3.18]{\text{\AA}}$.  The lattice constant is set
to $\unit[2.5]{\text{\AA}}$ for both lattices.  The calculations are
performed in the same way as described in the previous section with a
k-point sampling of ($45\times 45$) in the in-plane direction.  Also for this
system we have found that the PPA yields almost identical results to
the full frequency G$_0$W$_0$ and therefore all calculations presented
in this section have been performed with the PPA.

\begin{figure}[t]
\begin{centering}
\includegraphics[width=0.85\columnwidth,clip=]{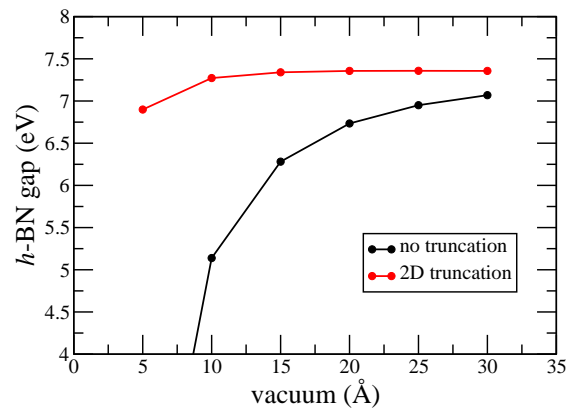}
\par\end{centering}
\caption{\label{fig:BN-vcut_gaps}(Color online.) Direct band gap at
  the K-point for a freestanding \textit{h}-BN sheet as function of
  the vacuum used to seperate layers in neighboring supercell with
  and without use of the Coulomb truncation method as described in
  Sec. \ref{subsec:truncation}.}
\end{figure}

The importance of truncating the Coulomb potential in order to avoid
spurious interaction between neighboring supercells is shown in
Fig. \ref{fig:BN-vcut_gaps} for the direct gap at the K-point for a
freestanding boron nitride monolayer. Without truncation, the gap converges 
very slowly with the cell size and is still $\unit[0.3]{eV}$ below the
converged value for $\unit[30]{\text{\AA}}$ of vacuum.  
Applying the truncated Coulomb potential, the band gap is clearly converged
already for $\unit[10]{\text{\AA}}$ vacuum. 
These observations are consistent with recent G$_0$W$_0$ calculations for a SiC
sheet, where the same trends were found.\cite{Hsueh_PRB2011}

\begin{figure}[t]
\begin{centering}
\includegraphics[width=0.85\columnwidth,clip=]{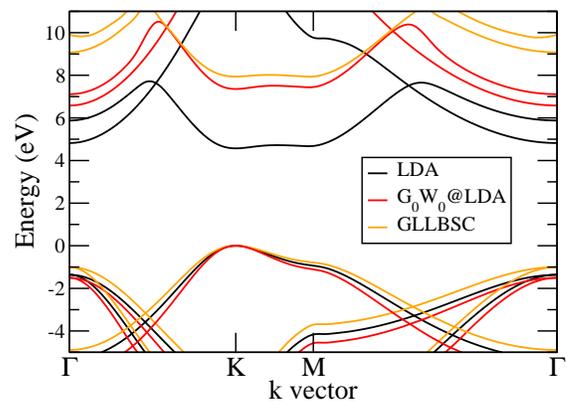}
\par\end{centering}
\caption{\label{fig:BN-bandstructure}(Color online.) Band structure for
  a freestanding \textit{h}-BN sheet. The band gap is direct at the
  K-point with LDA ($\unit[4.57]{eV}$) and GLLBSC ($\unit[7.94]{eV}$)
  and changes to indirect between the K- and the $\Gamma$-point for
  G$_0$W$_0$ ($\unit[7.37]{eV}$).}
\end{figure}

First, we summarize the most important features of the band structure
calculations for the freestanding \textit{h}-BN as shown in
Fig. \ref{fig:BN-bandstructure}.  LDA predicts a direct band gap at
the K-point of $\unit[4.57]{eV}$ and an indirect K-$\Gamma$ transition
of $\unit[4.82]{eV}$.  With GLLBSC, the bands are shifted
significantly in energy. However, the shift is not constant for the
different bands, resulting in a larger increase of the gap at the
$\Gamma$-point than at the K-point.  This yields $\unit[7.94]{eV}$ and
$\unit[9.08]{eV}$ for the direct and indirect transition,
respectively.  The opposite is the case for G$_0$W$_0$@LDA
calculations which predict an indirect band gap of $\unit[6.58]{eV}$
and a direct transition at the K-point of $\unit[7.37]{eV}$.  These
values are $\unit[0.6]{}$ and $\unit[1.0]{eV}$ larger than the ones
reported in Ref. \onlinecite{Blase_PRB1995} which were obtained from
pseudopotential-based G$_0$W$_0$@LDA calculations. We note, however,
that the amount of vacuum used in Ref. \onlinecite{Blase_PRB1995} was
only $\unit[13.5]{\text{\AA}}$ which is not sufficient according to
our results.

For the freestanding graphene (not shown), we find from the slope of
the Dirac cone at the K-point the Fermi velocity to be $\unitfrac[0.87
\cdot 10^6]{m}{s}$, $\unitfrac[0.87 \cdot 10^6]{m}{s}$ and
$\unitfrac[1.17 \cdot 10^6]{m}{s}$ with LDA, GLLBSC and G$_0$W$_0$,
respectively.  This is in good agreement with previous G$_0$W$_0$
calculations which obtained $\unitfrac[1.15 \cdot 10^6]{m}{s}$
(Ref. \onlinecite{Yang_PRL2009}) and $\unitfrac[1.12 \cdot
10^6]{m}{s}$ (Ref. \onlinecite{Trevisanutto_PRL2008}), respectively,
and accurate magnetotransport measurements which yielded
$\unitfrac[1.1 \cdot 10^6]{m}{s}$ (Ref. \onlinecite{Zhang_nature2005}).

\begin{figure}[t]
\begin{centering}
\includegraphics[width=0.85\columnwidth,clip=]{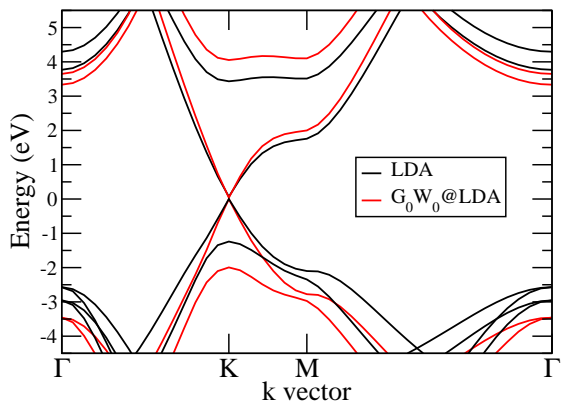}
\par\end{centering}
\caption{\label{fig:graphene-BN_bandstructure}(Color online.) LDA and
  G$_0$W$_0$@LDA band structure for a graphene/boron nitride double
  layer structure.  Only the two highest valence bands and the two
  lowest conduction bands are shown.}
\end{figure}

The band structure of graphene on a single \textit{h}-BN sheet is
shown in Fig. \ref{fig:graphene-BN_bandstructure}.  At a qualitative
level the band structure is similar to a superposition of the band
structures of the isolated systems.  In particular, due to the
limited coupling between the layers, the bands closest to the Fermi
energy can clearly be attributed to the different layers: At the
K-point, the linear dispersion of the graphene bands is maintained and
the second highest valence and second lowest conduction band belong to
the \textit{h}-BN.  However, there are important quantitative
changes. First, the slope of the Dirac cone is reduced, giving a Fermi
velocity of $\unitfrac[1.01 \cdot 10^6]{m}{s}$ ($\unitfrac[0.78 \cdot
10^6]{m}{s}$) with G$_0$W$_0$ (LDA).  Exactly at the K-point both LDA
and G$_0$W$_0$ predict a small gap of $\unit[50]{meV}$. Moreover, at
the K-point, the \textit{h}-BN gap obtained with G$_0$W$_0$ is reduced
from $\unit[7.37]{eV}$ for the isolated sheet to $\unit[6.35]{eV}$.
In contrast the LDA gap is almost the same ($\unit[4.67]{eV}$) as for
the isolated \textit{h}-BN.

\begin{figure}[t]
\begin{centering}
\includegraphics[width=0.85\columnwidth,clip=]{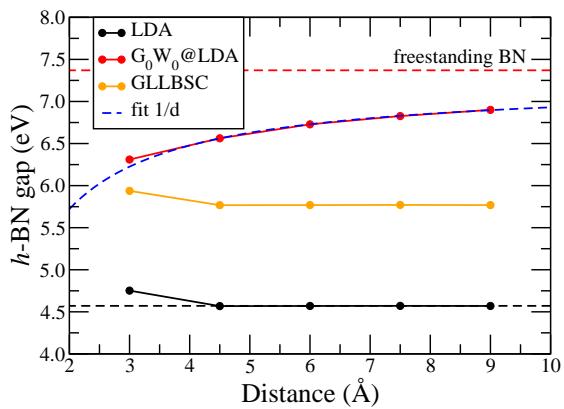}
\par\end{centering}
\caption{\label{fig:graphene-BN_distance}(Color online.) The band gap
  of \textit{h}-BN at the K-point as function of the distance to the
  graphene sheet (see inset).  Dashed horizontal lines indicate the
  values for the freestanding \textit{h}-BN, corresponding to $d
  \rightarrow \infty$.}
\end{figure}

To further illustrate the importance of screening effects, we
calculate the dependence of the \textit{h}-BN gap with respect to the
distance between the two layers. From Fig. \ref{fig:graphene-BN_distance},
we can see that for LDA the gap is almost constant at the value of the
freestanding boron nitride. For GLLBSC, the gap is around $\unit[1.2]{eV}$
larger but it does not change with the interlayer distance either. In contrast,
GW predicts an increase of the gap with increasing distance and slowly
approaches the value of the isolated system. The distance dependence of the
gap is well fitted by $1/d$ as expected from a simple image charge model. Only
for small distances, the results deviate from the $1/d$ dependence,
most likely due to the formation of a chemical bond between the layers.
We mention that the band gap closing due to substrate screening has
been observed in previous GW studies of metal/semiconductor
interfaces\cite{Charlesworth_PRL1993, Inkson_JPhysC1973} as well as
for molecules on metal surfaces.\cite{Neaton_PRL2006, Garcia_PRB2009,
  Thygesen_PRL2009, Freysoldt_PRL2009}
 
\begin{figure}[t]
\begin{centering}
\includegraphics[width=0.85\columnwidth,clip=]{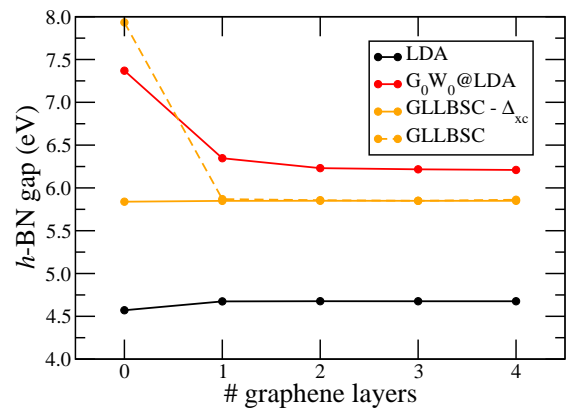}
\par\end{centering}
\caption{\label{fig:graphenex_gaps}(Color online.) \textit{h}-BN gap
  at the K-point for different number of adsorbed graphene layers.
  GLLBSC results are plotted without and with the derivative discontinuity
  $\Delta_{xc}$.}
\end{figure}

In Fig. \ref{fig:graphenex_gaps}, the size of the \textit{h}-BN gap is
shown for a varying number of graphene layers in a
\textit{h}-BN/$N$-graphene heterostructure.  While LDA predicts a
constant band gap of \textit{h}-BN, G$_0$W$_0$ predicts a slight
decrease of the gap with increasing number of graphene layers due to
enhanced screening. Additionally, we show the results for GLLBSC with
and without the derivative discontinuity $\Delta_{xc}$ added to
the Kohn-Sham gap. Due the construction of the GLLBSC,
$\Delta_{xc}$ vanishes when one or more graphene layers are
present because the system becomes (almost) metallic. Thus the GLLBSC
gap becomes independent of the number of graphene layers, but is still
close to the G$_0$W$_0$ result.

\section{Molecules}\label{sec:molecules}

In this section, we present G$_0$W$_0$ calculations for a set of 32
small molecules.  Recently a number of high-level GW studies on
molecular systems have been published.\cite{Rostgaard_PRB2010,
  Blase_PRB2011, Bruneval_JCTC2013, Caruso_PRB2012} These studies have
all been performed with localized basis sets and have explored the
consequences of many of the commonly made approximations related to
self-consistency, starting point-dependence in the G$_0$W$_0$
approach, and treatment of core electrons. Here we use the more
standard G$_0$W$_0$@LDA method and apply a plane wave basis set. This
is done in order to benchmark the accuracy of this scheme but also to
show the universality of the present implementation in terms of the
types of systems that can be treated.

\begin{figure}[t]
\begin{centering}
\includegraphics[width=0.85\columnwidth,clip=]{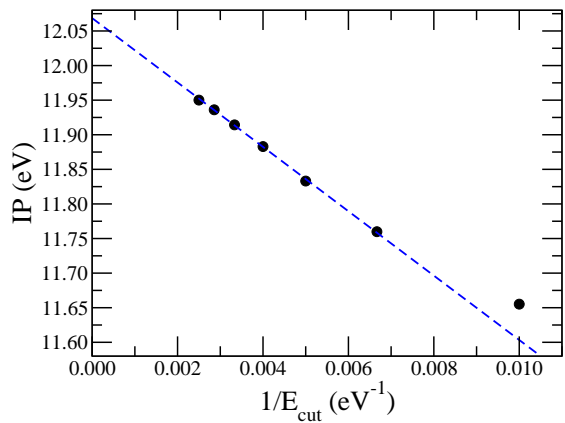}
\par\end{centering}
\caption{\label{fig:H2O}(Color online.) Convergence of the Ionization
  Potential for \sffamily H$_2$O \rmfamily with respect to the plane
  wave cutoff for G$_0$W$_0$@LDA.  The dashed line shows a linear fit
  of the points with $E_{\text{cut}} > \unit[100]{eV}$
  ($1/E_{\text{cut}} < \unit[0.01]{eV^{-1}}$). The IP is given as the
  negative HOMO energy.}
\end{figure}

Our calculations are performed in a supercell with
$\unit[7]{\text{\AA}}$ distance between neighboring molecules in all
directions.  As pointed out in the previous sections, careful
convergence tests are crucial in order to obtain accurate results with
GW.  For a plane wave basis we have found that this is particularly
important for molecules, as demonstrated in Fig. \ref{fig:H2O} for
water. Here, we plot the calculated ionization potential as a function
of the inverse plane wave cutoff.  Again, for each data point, the
number of bands is set equal to the number of plane waves
corresponding to the cutoff.  Even for $E_{\text{cut}} =
\unit[400]{eV}$ ($1/E_{\text{cut}} = \unit[0.0025]{eV^{-1}}$ and
corresponding to more than 8000 bands), the IP is not fully converged.
However, for a cutoff larger than $\unit[100]{eV}$, the IP grows
linearly with $1/E_{\text{cut}}$ and this allows us extrapolate to the
inifinite cutoff (and number of empty bands) limit.
\cite{Kang_PRB2010, Umari_PhysSol2011} In this case the converged
ionization potential is $\unit[12.1]{eV}$ which is about
$\unit[0.5]{eV}$ smaller than the experimental value.  For all the
molecules we have extrapolated the IP to infinite plane wave cutoff
based on G$_0$W$_0$ calculations at cutoff energies $\unit[200 -
400]{eV}$.  Furthermore, as found for the solids and the 2D systems,
the plasmon pole approximation and the fully frequency dependent GW
calculations yield very similar results with typically $0.05$ to
$\unit[0.1]{eV}$ smaller IPs for the latter.

\begin{figure}[t]
\begin{centering}
\includegraphics[width=0.85\columnwidth,clip=]{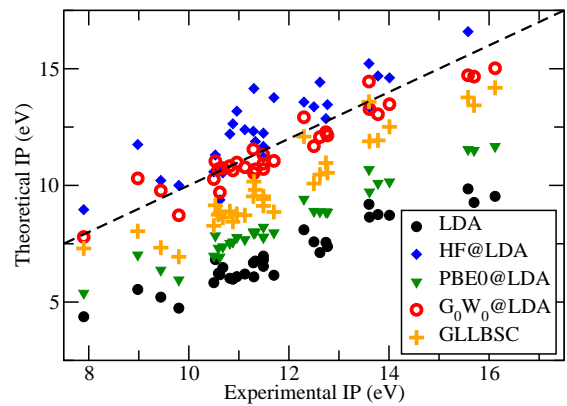}
\par\end{centering}
\caption{\label{fig:IPs}(Color online.) Comparison of theoretical and
  experimental ionization potentials.  The G$_0$W$_0$ results are
  obtained by applying the extrapolation scheme as explained in the
  text.  Corresponding values are listed in Tab. \ref{tab.num}.}
\end{figure}

The results for all molecules are summarized and compared in
Fig. \ref{fig:IPs}.  The LDA, PBE0 and GLLBSC calculations
underestimate the IP with mean absolute errors (MAE) of
$\unit[4.8]{}$, $\unit[3.5]{}$, and $\unit[2.0]{eV}$, respectively.
The opposite trend is observed for (non-selfconsistent) Hartree-Fock
which systematically overestimates the IP due to complete lack of
screening.  The MAE found for HF is $\unit[1.1]{eV}$.  We note that
for an exact functional, according to the ionization-potential
theorem, the Kohn-Sham energy of the highest occupied molecular
orbital (HOMO) from DFT should be equal to the negative ionization
potential.\cite{Perdew_PRL1982}

\begin{figure}[t]
\begin{centering}
\includegraphics[width=0.85\columnwidth,clip=]{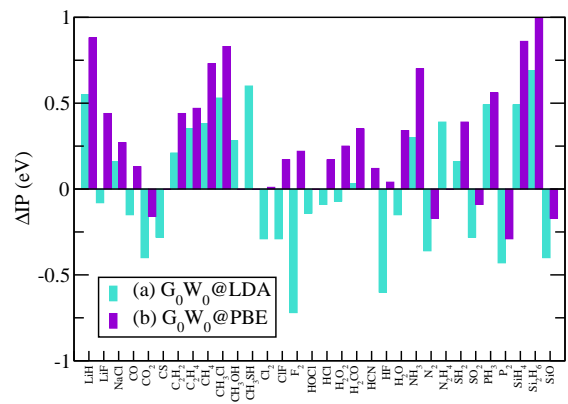}
\par\end{centering}
\caption{\label{fig:IPdiff}(Color online.) Deviations for the
  ionization potentials obtained with G$_0$W$_0$@LDA compared to (a)
  Ref. \onlinecite{Bruneval_JCTC2013} and (b)
  Ref. \onlinecite{Caruso_PRB2012}. The mean deviations are
  $\unit[0.02]{}$ and $\unit[0.30]{eV}$, respectively.}
\end{figure}

The G$_0$W$_0$ results are typically around $\unit[0.5]{eV}$ smaller
than the experimental IPs, although there are a few exceptions where
the calculated ionization potential is too large, and with a MAE of
$\unit[0.56]{eV}$.  Recently, very similar studies have been reported
for G$_0$W$_0$@LDA\cite{Bruneval_JCTC2013} with Gaussian basis sets
and G$_0$W$_0$@PBE\cite{Caruso_PRB2012} in an all-electron framework
using numerical atomic orbitals. Although there are differences of up
to $\unit[0.5]{eV}$ (both positive and negative), we find reasonable
overall agreement with $\unit[0.32]{eV}$ MAE relative to
Ref. \onlinecite{Bruneval_JCTC2013}. The mean signed error (MSE) is
only $\unit[0.02]{eV}$.
Compared to Ref. \onlinecite{Caruso_PRB2012}, our results are
systematically smaller with a MAE of $\unit[0.36]{eV}$ and a MSE of
$\unit[0.30]{eV}$. This is within the range of the accuracy of the
different implementations, e.g. basis set, the PPA and the frozen
core approximation applied in our calculations and the
differences between LDA and PBE as starting points. A graphical
comparison with these studies is shown in Fig. \ref{fig:IPdiff}.

For detailed discussions of the role of self-consistency and other
approximations we refer to Refs.  \onlinecite{Rostgaard_PRB2010,
  Caruso_PRB2012, Stan_JChemPhys2009, Bruneval_JCTC2013}.

\begin{table*}[tp]
\begin{centering}
  \caption{Calculated and experimental ionization potentials. All
    energies are in eV. Last row shows the mean absolute error (MAE)
    with respect to experiments. Experimental data taken from
    Ref. \onlinecite{nist}.}\label{tab.num}
\begin{tabularx}{0.7\textwidth}{X *{6}{r@{.}l}}
\hline\hline
  Molecule       & \multicolumn{2}{c}{LDA} & \multicolumn{2}{c}{HF@LDA} &
\multicolumn{2}{c}{PBE0@LDA} & \multicolumn{2}{c}{GLLBSC} &
\multicolumn{2}{c}{G$_0$W$_0$@LDA} &
\multicolumn{2}{c}{experiment}\\
  \hline
  LiH         &
\hspace{6mm}4&37\hspace{4mm} & \hspace{6mm}8&96\hspace{4mm} &
\hspace{6mm}5&38\hspace{4mm} & \hspace{6mm}7&30\hspace{4mm} &
\hspace{6mm}7&79\hspace{4mm} & \hspace{6mm}7&90\hspace{4mm} \\
LiF		&	6&08	&	14&15	&	7&95	&	10&16
&	10&53	&	11&30	\\
NaCl		&	4&74	&	10&00	&	5&95	&	6&94
&	8&72	&	9&80	\\
CO		&	8&72	&	14&61	&	10&15	&	12&51
&	13&48	&	14&01	\\
CO$_2$		&	8&75	&	14&69	&	10&09	&	11&93
&	13&05	&	13&78	\\
CS		&	6&76	&	11&88	&	8&00	&	9&81
&	10&69	&	11&33	\\
C$_2$H$_2$	&	6&81	&	11&21	&	7&79	&	9&41
&	11&22	&	11&49	\\
C$_2$H$_4$	&	6&48	&	10&54	&	7&37	&	8&62
&	10&74	&	10&68	\\
CH$_4$		&	9&19	&	15&22	&	10&68	&	13&58
&	14&45	&	13&60	\\
CH$_3$Cl	&	6&68	&	12&32	&	8&01	&	9&53
&	11&55	&	11&29	\\
CH$_3$OH	&	6&09	&	13&18	&	7&77	&	8&77
&	10&98	&	10&96	\\
CH$_3$SH	&	5&21	&	10&21	&	6&37	&	7&33
&	9&78	&	9&44	\\
Cl$_2$		&	6&53	&	11&67	&	7&77	&	9&12
&	10&93	&	11&49	\\
ClF		&	7&38	&	13&46	&	8&85	&	10&54
&	12&14	&	12&77	\\
F$_2$		&	9&27	&	18&44	&	11&50	&	13&43
&	14&66	&	15&70	\\
HOCl		&	6&20	&	12&39	&	7&68	&	8&72
&	10&78	&	11&12	\\
HCl		&	7&56	&	12&86	&	8&87	&	10&96
&	12&28	&	12&74	\\
H$_2$O$_2$	&	6&15	&	13&76	&	7&97	&	8&86
&	11&05	&	11&70	\\
H$_2$CO		&	5&98	&	12&64	&	7&58	&	8&44
&	10&64	&	10&88	\\
HCN		&	8&64	&	13&35	&	9&72	&	11&89
&	13&27	&	13&61	\\
HF		&	9&53	&	18&29	&	11&67	&	14&18
&	15&02	&	16&12	\\
H$_2$O		&	7&12	&	14&42	&	8&87	&	10&46
&	12&07	&	12&62	\\
NH$_3$		&	6&02	&	12&20	&	7&52	&	8&89
&	10&83	&	10&82	\\
N$_2$		&	9&85	&	16&59	&	11&54	&	13&77
&	14&72	&	15&58	\\
N$_2$H$_4$	&	5&54	&	11&75	&	7&02	&	8&04
&	10&30	&	8&98	\\
SH$_2$		&	5&83	&	10&58	&	6&97	&	8&27
&	10&27	&	10&50	\\
SO$_2$		&	7&58	&	13&37	&	8&89	&	10&08
&	11&68	&	12&50	\\
PH$_3$		&	6&23	&	10&77	&	7&31	&	8&74
&	10&70	&	10&59	\\
P$_2$		&	6&17	&	9&38	&	6&93	&	8&80
&	9&70	&	10&62	\\
SiH$_4$		&	8&10	&	13&57	&	9&41	&	12&09
&	12&92	&	12&30	\\
Si$_2$H$_6$	&	6&82	&	11&30	&	7&84	&	9&15
&	11&04	&	10&53	\\
SiO		&	6&97	&	12&24	&	8&21	&	9&53
&	10&70	&	11&49	\\
\hline
MAE		&	4&84 	& 	1&11 	& 	3&46 	& 	1&83 
& 	0&56 	&	  &	\\
\hline\hline
\end{tabularx}
\end{centering}
\end{table*}

\section{Conclusions}

We have presented a plane-wave implementation of the single-shot G$_0$W$_0$
approximation within the GPAW projector augmeted wave method code.
The method has been applied to the calculation of quasiparticle band
structures and energy levels in bulk crystals, 2D materials, and
molecules, respectively.  Particular attention has been paid to the
convergence of the calculations with respect to the plane wave cutoff
and the number of unoccupied bands. While for all extended systems the
value of the band gap was found to be converged at around
$\unit[200]{eV}$, the ionization potentials of the molecules required
significantly higher cutoffs. In these cases, the data points were
fit linearly to $1/E_{\text{cut}}$, allowing to extrapolate to
infinite number of bands. For all calculations, the plasmon pole
approximation and the use of full frequency dependence of the
dielectric function and the screened potential give very similar
results. With these two observations, the computational demands can be
drastically reduced without losing accuracy.

For the bulk semiconductors, we found good agreement with experimental
results with a mean absolute error (MAE) of $\unit[0.2]{eV}$.  However, in the
special case of zinc oxide and for the large gap insulators, the
calculated band gaps were underestimated by $\unit[0.5 - 1]{eV}$.
These errors are most likely due to the lack of self-consistency
and/or the quality of the LDA starting point used in our calculations.
Similar conclusions apply to the 32 small molecules where the
ionization potentials obtained from G$_0$W$_0$@LDA were found to
underestimate the experimental values by around \unit[0.5]{eV}
on average. The important role of screening for the
quasiparticle band structure was illustrated by the case of a
2D graphene/boron-nitride heterojunction. For this system, we found
a truncation of the Coulomb potential to be crucial in periodic
supercell calculations.

The G$_0$W$_0$ results were compared to band structures obtained with
Hartree-Fock, the PBE0 hybrid and the GLLBSC potential.  While
Hartree-Fock and PBE0 yield overall poor results, the computationally
efficient GLLBSC results were found to be in surprisingly good
agreement with G$_0$W$_0$ for the band gaps of semiconductors, while
the ionization potentials of molecules were found to be
$\unit[1.5]{eV}$ lower on average.

\section*{Acknowledgements}

We would like to thank Jun Yan and Jens J{\o}rgen Mortensen for useful 
discussions and assistance with the coding.
The authors acknowledge support from the Danish Council for Independent
Research's Sapere Aude Program through grant no. 11-1051390.
The Center for Nanostructured Graphene is sponsored by the Danish
National Research Foundation.
The Catalysis for Sustainable Energy (CASE) initiative is
funded by the Danish Ministry of Science, Technology and Innovation.

\bibliographystyle{apsrev}

\end{document}